\documentclass[a4paper,10pt]{article}
\usepackage[utf8x]{inputenc}

\title{ $M$-Theory on  Deformed Superspace }
\author{Mir Faizal\\ Department of Mathematics,  Durham University,\\ Durham, DH1 3LE,  United Kingdom,\\ faizal.mir@durham.ac.uk}

\begin{document}

\maketitle

\begin{abstract}
In this paper we will analyse a   noncommutative deformation of the 
ABJM theory in $N=1$ superspace formalism. 
We will then analyse the BRST 
and the anti-BRST symmetries for  this deformed  ABJM  theory, in linear as well as non-linear gauges.
  We will show that 
the sum of gauge fixing term and the ghost term  for this deformed  ABJM  theory can be expressed 
as a combination of  the total BRST and the total anti-BRST variation, in   Landau  and  non-linear gauges. 
We will show that in Landau  and  Curci-Ferrari gauges this deformed ABJM theory is invariant under a 
additional set of symmetry transformations. We will also discuss the effect that the  addition of a  bare mass term has on this 
theory. 
\end{abstract}

Key words:   ABJM Theory, BRST, Anti-BRST, Deformed Superspace    

PACS number: 11.25.Yb

\section{Introduction}
The construction of a action for $M$-theory  at low 
energies with manifest $N =8$ superconformal symmetry has led to the 
discovery of the  Bagger and Lambert  action with a Lie $3$-algebra \cite{1}-\cite{5}.  However,
 only one example of 
such a such $3$-algebra is known  and  so far the   rank of the gauge group has not been increased.
But a  $U(N)_k \times U(N)_{-k}$ superconformal Chern-Simons-matter 
 theory with level $k$ and $-k$ with arbitrary rank and   
$N = 6$ supersymmetry has been constructed \cite{apjm}. 
This theory called the ABJM theory is thought to describe the 
the world-volume of $N$ $M2$-membranes 
placed at the singularity of $R^8/Z_k$. This is because
 it may be possible to  
enhanced the supersymmetry of this theory to $N = 8$ supersymmetry \cite{su}.
 Furthermore, if this is done then 
 a $SO(8)$ $R$-symmetry at Chern-Simons levels
$k = 1,2$ will also exists for this theory. 

Chern-Simons theory in $N = 1$ superspace  formalism  has also been used in analysing the 
low energy approximation of the  action for  $M$-theory with $N =8$ supersymmetry \cite{14}. 
This was done by constructing a  manifestly $SO(7)$ invariant super-potential which  for specially chosen couplings, reproduced the 
 Bagger and Lambert  action \cite{2,3}. Hence for these values of the coupling constants  full $SO(8)$ symmetry was restored.
Chern-Simons theory in $N = 1$ superspace formalism has also been used for studding  the ABJM theory \cite{ab1}. 
By using the  Higgs mechanism    
higher-order terms that occur in the low energy approximation of the  action for  $M$-theory  have  been  analysed
in this $N =1$ superspace formalism.

The presence of a $NS$ antisymmetric tensor background
is a source of spacetime noncommutativity in string theory
 \cite{sw, dn}.
Now as string theory  introduces noncommutativity in spacetime, so 
field theories with spacetime noncommutativity have been thoroughly studied \cite{dfr}-\cite{av}.
The extension of spacetime noncommutativity to superspace noncommutativity
is related to the presence of other background fields. The
$RR$ field strength backgrounds  give rise to 
$\theta$-$\theta$ type deformations \cite{ov,se}
and   a gravitino background give rise to  $x$-$\theta$ deformation \cite{bgn}.
As superspace  noncommutativity  also arises in string theory, 
field theories with superspace noncommutativity have also been thoroughly studied \cite{se}-\cite{bs}. 
However, the presence of $\theta$-$\theta$ 
deformation breaks half of the supersymmetry. As we want to retain 
all the supersymmetry in our theory, we 
 not include  the $\theta$-$\theta$ deformation 
of the ABJM theory, in this paper.
It may be noted that even though this is the
 first work on noncommutative deformation of the ABJM theory, we 
analyse both the $x$-$x$ and $x$-$\theta$ deformations at the same time.
 This because  we use a similar
 formulism to analyse both these deformations. 

 Due to the duality between $M$-theory and $II$ string theory, 
we expect that a 
 deformation of the super algebra on the string theory side will also
correspond to some deformation of the super algebra on the $M$-theory side.  
It is interesting  to note that a three-form field strength occurs
 naturally in $M$-theory. 
Besides that $M2$-branes in $M$-theory can end on $M5$-branes. 
In this sense $M5$ branes 
  in $M$-theory act as analogous objects 
to a $D$-brane in string theory. So we expect that coupling 
the ABJM theory to a 
background three-form field could lead to a noncommutative 
deformation of its super algebra, 
just like a background two-form  field 
strength leads to  a  noncommutative deformation of the super albegra of  
$D$-branes.  
This can be useful in describing the physics of $M2$-branes 
ending on $M5$-branes. 
 It may be noted that action for a single 
$M5$-brane  can be derived by demanding
the $\kappa$-symmetry of the open membrane ending on it \cite{14om}
 Thus the analysis of 
ABJM theory coupled to a background three-form field strength might 
give some useful insights into understanding the dynamics of multiple 
$M5$-branes. 
This will be interesting  because  even though the action for a single
 $M5$-brane is known, the  
action   for multiple $M5$-branes is not
known \cite{41s}-\cite{42s}.

We will thus analyse the noncommutative
  deformations of the ABJM theory that is expected to occur due to the  
coupling of the ABJM theory with  the background 
three-form field strength that occurs
 naturally in the $M$-theory.
  As the ABJM theory is composed of two 
Chern-Simons theories suitably coupled to matter fields, 
 a Seiberg-Witten map will hold for the noncommutative ABJM theory because 
it is known to hold for noncommutative Chern-Simons theories
 \cite{cssw, cssw1}. 
We thus analyse this noncommutative 
ABJM theory by relating the noncommutative
 fields in it to ordinary commutative fields.
 The product of these noncommutative fields 
will then induces a star product for the ordinary commutative fields.

The BRST  and the anti-BRST symmetries for gauge theories have been
 thoroughly studied \cite{abcd}. 
In fact it is known that for the Yang-Mills theories  in Landau 
 and non-linear gauges the algebra generated by the BRST
and the anti-BRST transformations along with $FP$-conjugation is a
 sub-algebra of a larger algebra called  Nakanishi-Ojima  algebra
\cite{n01}-\cite{n04}. The effect of addition of a bare mass term 
on the BRST and the anti-BRST symmetries has also been analysed in the 
non-linear gauges \cite{n001}.
The BRST symmetry for the Chern-Simons theory has also been
 thoroughly investigated
\cite{16,17}. The BRST symmetry of $N = 1$  abelian   
Chern-Simons theory \cite{18} and $N = 1$ non-abelian
Chern-Simons theory \cite{19} has been analysed in the
 superspace formalism. 
The BRST symmetry  of noncommutative pure Chern-Simons theory 
has also been analysed \cite{20g,21g}.
We will analyse the BRST and the anti-BRST symmetries of this 
deformed ABJM theory. The main focus of this paper will be the
 generalization of some known results about 
 the BRST and the anti-BRST 
symmetries in Yang-Mills theories to this deformed ABJM theory. 
 In particular we will show that, in certain gauges, the sum of this 
deformed ABJM theory  along with a gauge fixing term and a ghost term
 is invariant under a set of symmetry transformations which obey 
$ SL(2, R)$ algebra. 
This  is a similar to the invariance of Yang-Mills theories  under the 
 Nakanishi-Ojima Algebra \cite{n01}. 
 Furthermore,  it is known that the evolution 
of the $S$-matrix in the  Yang-Mills theories
 in massive Curci-Ferrari gauge is not unitary because  
the bare mass term breaks the nilpotency
 of the BRST and the anti-BRST transformations \cite{n001}.   
 We will show that a similar result 
holds for this deformed ABJM theory in massive Curci-Ferrari gauge.
Thus we will show that for ABJM theory 
the unitarity of $S$-matrix is violated
in massive
 Curci-Ferrari gauge due to the breaking of nilpotency of the 
BRST and the anti-BRST transformations.

\section{Deformation of ABJM Theory }
In this section we will deform the superspace of ABJM theory
 without breaking any supersymmetry. 
To do so we define $\theta^a$ as a two-component Grassmann parameter and let $ y^\mu = x^\mu + \theta^a (\gamma^\mu) _{a}^b \theta_b$. 
Then we promote them to operators $\hat{\theta}^a$ and $\hat{y}^\mu$ such that they satisfy the following
 deformed superspace algebra \cite{bgn}, 
\begin{eqnarray}
 [\hat{y}^\mu, \hat{y}^\nu] =  B^{\mu\nu}, && 
 {[\hat{y}^\mu, \hat{\theta}^a]} = A^{\mu a }. 
\end{eqnarray}
This is the most general deformation that we can have without breaking any supersymmetry \cite{se}. 
We use Weyl
ordering and  express the Fourier transformation of a superfields on this deformed superspace, as  
\begin{equation}
\hat{X} (\hat{y}, \hat{\theta}) =
\int d^3 k \int d^2 \pi \, \, e^{-i k \hat{y} -\pi \hat{\theta} } \,
X(k,\pi).
\end{equation}
Now  we  have a one to one map between a function of
$\hat{\theta}, \hat{y}$ to a function of ordinary
 superspace coordinates $\theta, y$ via
\begin{equation}
X (y, \theta)  =
\int d^3 k \int d^2 \pi \, \, e^{-i k y -\pi \theta } \,
X (k,\pi).
\end{equation}
Now as we have a one to one map between superfields  on this deformed superspace with 
superfields on the undeformed superspace, we can define   the product 
of two superfields on this deformed superspace. To do that
 we can express the product of two superfields  
${\hat{X}}(\hat{y},\hat{\theta}) { \hat{Z} } (\hat{y},\hat{\theta})$
on this deformed superspace, as
\begin{eqnarray}
{\hat{X}}(\hat{y},\hat{\theta}) { \hat{Z}}  (\hat{y},\hat{\theta}) &=&
\int d^3 k_1 d^3 k_2 \int d^2 \pi_1  d^2 \pi_2\, \, 
\exp -i( ( k_1 +k_2) \hat{y} +(\pi_1 +\pi_2) \hat{\theta} )  \nonumber \\ && 
\,\,\,\,\,\, \,\,\,\,\,\, \,\,\,\,\,\, \,\,\,\, \,\,\,\,\,\, \,\,\,\, \,\,\,\,\,\, \,\,\,\, \,\,\,\,\,\, \times  \exp(i\Delta)
\,{X}  (k_1,\pi_1) {Z}   (k_2,\pi_2),
\end{eqnarray}
where
\begin{equation}
\exp (i\Delta) = \exp -\frac{i}{2} \left(  B^{\mu\nu} k^2_\mu k^1_\nu
+  A^{\mu a} (\pi^2 _a k^1_\mu - k^2_\mu \pi^1_a \right).
\end{equation}
So we can now define the star product  between ordinary functions  as follows:
\begin{eqnarray}
{X}(y,\theta) \star { Z}  (y,\theta) & =& 
\exp -\frac{i}{2} \left(
 B^{\mu\nu} \partial^2_\mu \partial^1_\nu
+ A^{\mu a} (\partial^2 _a \partial^1_\mu - \partial^2_\mu \partial^1_a \right)) \nonumber \\ &&
\,\,\,\,\,\, \,\,\,\,\,\, \,\,\,\,\,\,\,\,\,\,
 \times 
 {X}(y_1,\theta_1) {Z}  (y_2, \theta_2)
\left. \right|_{y_1=y_2=y, \; \theta_1=\theta_2=\theta}.
\label{star2}
\end{eqnarray}
The star product reduces to the usual Moyal star product
for the bosonic noncommutativity  in the limit $A^{a\mu}=0$. 
Furthermore, when both $B^{\mu\nu} = A^{a\mu}=0$, then the star product 
reduces to the ordinary product.  
It is also useful to define the following bracket 
\begin{equation}
 2[X, Z]_{\star} =  X \star Z \pm  Z \star X, 
\end{equation}
where the relative sign is negative unless  both the fields are fermionic.

Now we construct the  classical Lagrangian density with the gauge group $U(N)_k \times U(N)_{-k}$ \cite{ab1},
  on this deformed superspace, 
\begin{equation}
{ \mathcal{L}_c} =  \mathcal{L}_{M} + \mathcal{L}_{CS} - \tilde{\mathcal{L}}_{CS},
\end{equation}
where $\mathcal{L}_{CS}$ and  $\tilde{\mathcal{L}}_{CS}$ are deformed  Chern-Simons theories with gauge group's $U(N)_k$ and $U(N)_{-k}$
respectively. They   can thus be expressed as  
\begin{eqnarray}
 \mathcal{L}_{CS} &=& \frac{k}{2\pi} \int d^2 \,  \theta \, \, 
  Tr \left[  \Gamma^a  \star  \omega_a + \frac{i}{3} [\Gamma^a, \Gamma^b]_{\star}
\star D_b \Gamma_a \right.
\nonumber \\ &&\left. \,\,\,\,\,\, \,\,\,\,\,\, \,
\,\,\,\,\,\, \,\,\,\,\,\, \,\,\,\,\,\,\,\,\,\,\,
 + \frac{1}{3} [\Gamma^a,\Gamma^b]_{\star}\star[\Gamma_a, \Gamma_b]_{\star}
\right] _|, 
\nonumber \\
 \tilde{\mathcal{L}}_{CS} &=& \frac{k}{2\pi} \int d^2 \,  \theta \, \, 
  Tr \left[  \tilde{\Gamma}^a  \star  \tilde{\omega}_a
 + \frac{i}{3} [\tilde{\Gamma}^a, \tilde{\Gamma}^b]_{\star}
\star D_b \tilde{\Gamma}_a \right.
\nonumber \\ && \left. \,\,\,\,\,\,
 \,\,\,\,\,\, \,\,\,\,\,\,\, \,\,\,\,\,\, \,\,\,\,\,\,\,\,\,\,\,
+ \frac{1}{3} [\tilde{\Gamma}^a,\tilde{\Gamma}^b]_{\star}\star
[\tilde{\Gamma}_a, \tilde{\Gamma}_b]_{\star}\right] _|, 
\end{eqnarray}
where $k$ is an integer \cite{k} and 
\begin{eqnarray}
 \omega_a &=&  \frac{1}{2} D^b D_a \Gamma_b - i  [\Gamma^b, D_b \Gamma_a]_{\star } 
- \frac{2}{3} [ \Gamma^b ,
[ \Gamma_b, \Gamma_a]_{\star}]_{\star}\nonumber \\
   \tilde\omega_a &=&  \frac{1}{2} D^b D_a \tilde\Gamma_b -i  [\tilde\Gamma^b, D_b \tilde\Gamma_a]_{\star } - 
\frac{2}{3} [ \tilde\Gamma^b,
[ \tilde\Gamma_b,  \tilde\Gamma_a]_{\star } ]_{\star }. 
\end{eqnarray}
Here the super-derivative $D_a$ is given by 
\begin{equation}
 D_a = \partial_a + (\gamma^\mu \partial_\mu)^b_a \theta_b,
\end{equation}
and  $'|'$  means that the quantity is evaluated at $\theta_a =0$. 
In component form the $\Gamma_a$ and $\tilde \Gamma_a$ are given by 
\begin{eqnarray}
 \Gamma_a = \chi_a + B \theta_a + \frac{1}{2}(\gamma^\mu)_a A_\mu + i\theta^2 \left[\lambda_a -
 \frac{1}{2}(\gamma^\mu \partial_\mu \chi)_a\right], \nonumber \\
 \tilde\Gamma_a = \tilde\chi_a + \tilde B \theta_a + \frac{1}{2}(\gamma^\mu)_a \tilde A_\mu + i\theta^2 \left[\tilde \lambda_a -
 \frac{1}{2}(\gamma^\mu \partial_\mu \tilde\chi)_a\right]. 
\end{eqnarray}
The Lagrangian density for the matter fields  is given by 
\begin{eqnarray}
 \mathcal{L}_{M} &=& \frac{1}{4} \int d^2 \,  \theta \, \,  
Tr \left[ [\nabla^a_{(X)}  \star        X^{I \dagger}     \star     
\nabla_{a (X)}  \star        X_I ]\right. \nonumber \\ && \left. 
\,\,\,\,\,\, \,\,\,\,\,\, \,\,\,\,\,\,\,\,\,\, +
[\nabla^a_{(Y)} \star         Y^{I \dagger}  \star        \nabla_{a (Y)}    \star       Y_I ] + 
 \frac{16\pi}{k} \mathcal{V}_{\star   } \right]_|,
\end{eqnarray}
where 
\begin{eqnarray}
 \nabla_{(X)a}    \star     X^{I } &=& 
D_a  X^{I } + i \Gamma_a   \star      X^I 
-     i   X^I\star   \tilde\Gamma_a  , \nonumber \\ 
 \nabla_{(X)a}   \star      X^{I \dagger} 
&=& D_a  X^{I  \dagger}      
+ i \tilde\Gamma_a   \star    X^{I  \dagger}- 
     i X^{I  \dagger}\star  \Gamma_a, \nonumber \\ 
 \nabla_{(Y)a}    \star     Y^{I } &=& D_a  Y^{I }  
+ i \tilde\Gamma_a   \star      Y^I- i  Y^I \star \Gamma_a , \nonumber \\ 
 \nabla_{(Y)a}     \star    Y^{I \dagger} &=& D_a  Y^{I  \dagger} 
+ i \Gamma_a   \star 
       Y^{I  \dagger} 
 - i   Y^{I  \dagger}\star    \tilde\Gamma_a,
\end{eqnarray}
and $\mathcal{V}_\star $ is the potential term  given by 
\begin{equation}
 \mathcal{V}_\star  = \epsilon^{IJ} \epsilon_{KL} [ X_I\star  Y^K\star  X_J \star Y^L] + \epsilon_{IJ} \epsilon^{KL} 
[ X^{\dagger I}\star  Y^{ \dagger}_{ K}\star  X^{ \dagger J} \star Y^{ \dagger}_{ L} ]. 
\end{equation}
 This model reduces to the regular ABJM theory when $ B^{\mu\nu} = A^{\mu a } =0$.

\section{Linear Gauge}
All the degree's of freedom in the Lagrangian density for
 this  deformed  ABJM  theory are not physical because it is invariant
 under the following gauge transformations, 
\begin{eqnarray}
 \delta\, \Gamma_a = \nabla_a \star \Lambda, &&   \delta\, \tilde\Gamma_a
 = \tilde\nabla_a \star \tilde\Lambda, \nonumber \\ 
\delta\, X^{I } = i(\Lambda \star X^{I } -X^{I } \star  \tilde \Lambda ), 
 &&  \delta \, X^{I \dagger  } = i(\tilde \Lambda\star  X^{I\dagger  }
 -  X^{I\dagger  } \star \Lambda ),
\nonumber \\
\delta\, Y^{I } = i(  \tilde \Lambda \star Y^{I } -  Y^{I }\star \Lambda),  &&  
\delta \, Y^{I \dagger  } = i(\Lambda \star  Y^{I\dagger  }
-   Y^{I\dagger  }\star  \tilde \Lambda),
\end{eqnarray}
where 
\begin{eqnarray}
 \nabla_a = D_a -i \Gamma_a, && \tilde{\nabla}_a  = D_a - i \tilde{\Gamma}_a. 
\end{eqnarray}
So we have to fix a gauge before doing any 
calculations. This can be done by choosing the following   gauge fixing conditions,
\begin{eqnarray}
D^a \star \Gamma_a =0, && D^a  \star \tilde{\Gamma}_a =0. 
\end{eqnarray}
These  gauge fixing conditions can be incorporate at the quantum level by adding the following  gauge fixing term to 
the original Lagrangian density,
\begin{equation}
\mathcal{L}_{gf} = \int d^2 \,  \theta \, \, Tr  \left[b \star (D^a \Gamma_a) + \frac{\alpha}{2}b \star b -
i\tilde{b}  \star (D^a \tilde{\Gamma}_a) + \frac{\alpha}{2}\tilde{b}  \star \tilde b 
\right]_|.
\end{equation}
The ghost terms corresponding to this gauge fixing term can be written as  
\begin{equation}
\mathcal{L}_{gh} = \int d^2 \,  \theta \, \,  Tr 
[ \overline{c}  \star D^a \nabla_a  \star c - \tilde{\overline{c}}  \star D^a \tilde{\nabla}_a \star \tilde{c} ]_|.
\end{equation}
The total Lagrangian density obtained by addition of the original classical Lagrangian density, the gauge 
fixing term and the ghost term is invariant under the following BRST transformations, 
\begin{eqnarray}
s \,\Gamma_{a} = \nabla_a \star  c, && s\, \tilde\Gamma_{a} =\tilde\nabla_a  \star  \tilde c, \nonumber \\
s \,c = - {[c,c]}_ {\star} , && s \,\tilde{\overline{c}} =- \tilde b - 2 [\tilde{\overline{c}} ,  \tilde c]_{\star}, \nonumber \\
s \,\overline{c} = b, && s \,\tilde c = - [\tilde c, \tilde c]_{\star}, \nonumber \\ 
s \,b =0, &&s \, \tilde b= - [ \tilde b, \tilde{\overline{c}}]_{\star}, \nonumber \\ 
s \, X^{I } = i(c \star X^{I } -  X^{I }\star \tilde c), 
 &&  s \, X^{I \dagger }
 = i(  \tilde c \star  X^{I \dagger } -   X^{I \dagger }\star c ), 
\nonumber \\
s \, Y^{I } = i(  \tilde c \star Y^{I } -Y^{I }\star  c ),  &&  
s \, Y^{I \dagger } = i(c\star  Y^{I \dagger } -   Y^{I \dagger }
\star \tilde c). 
\end{eqnarray}
This total Lagrangian density is also invariant under the following anti-BRST transformations,
\begin{eqnarray}
\overline{s} \,\Gamma_{a} = \nabla_a  \star \overline{c}, &&  \overline{s} 
\, \tilde \Gamma_{a} =  \tilde\nabla_a \star \tilde{\overline{c}},\nonumber \\
\overline{s} \,c = -b - 2 [\overline{c}, c]_ \star,  && [\overline{s} ,\tilde c]_ \star =  \tilde b, \nonumber \\
\overline{s} \,\overline{c} = - [\overline{c}, \overline{c}]_\star, 
                   &&\overline{s} \,\tilde{\overline{c}} = -[ \tilde{\overline{c}},\tilde{\overline{c}}]_\star,\nonumber \\ 
\overline{s} \,b =- {[b,c]}_\star,  && \overline{s}  \,\tilde b  = 0,
 \nonumber \\ 
\overline{s} \, X^{I } = i(\overline{c}\star X^{I } -  X^{I }\star \tilde{\overline{c}}), 
&&   \overline{s} \, X^{I \dagger } = 
 i( \tilde{\overline{c}}\star  X^{I \dagger } -   X^{I \dagger }\star\overline{c}),\nonumber \\
\overline{s} \, Y^{I } = i( \tilde{\overline{c}}\star Y^{I }
 -  Y^{I }\star \overline{c}), 
&&   \overline{s} \, Y^{I \dagger } =  i(\overline{c}\star  Y^{I \dagger }
 -   Y^{I \dagger } \star \tilde{\overline{c}}).
\end{eqnarray}
Both these sets of transformations are nilpotent. 
\begin{equation}
[s, s]_\star = [\overline{s}, \overline{s}]_\star  = 0. 
\end{equation}
In fact they also satisfy $[s, \overline{s}]_\star = 0$. Here star product means that any product of fields in 
the transformation be treated as a star product. 
We can now express the  sum of the gauge fixing term and the ghost term as 
\begin{eqnarray}
\mathcal{L}_{gf} + \mathcal{L}_{gh} 
 &=&- \int d^2 \,  \theta \, \, \overline{s}\, Tr  \left[ c \star \left(D^a \Gamma_a   
 -  \frac{i\alpha}{2}b\right) - \tilde c \star \left(D^a \tilde\Gamma_a   
 -  \frac{i\alpha}{2}\tilde b\right) \right]_|\nonumber\\
 &=&  \int d^2 \,  \theta \, \,  s\, Tr \left[ \overline{c} \star \left(D^a  \Gamma_a 
 -  \frac{ \alpha}{2}b\right)
 -   \tilde{\overline{c}} \star \left(D^a  \tilde \Gamma_a  
 -  \frac{ \alpha}{2}\tilde b\right)
\right]_|.
\end{eqnarray}
Thus the sum of gauge fixing term and ghost term can be expressed as a total BRST or a total anti-BRST variation. 
In  Landau gauge, $\alpha = 0 $,  and so we have 
\begin{eqnarray}
\mathcal{L}_{gh} + \mathcal{L}_{gf}  &=&\int d^2 \,  \theta \, \,  s\,   Tr \left[ \overline{c} \star (D^a \Gamma_a  )  
- \tilde{\overline{c}} \star(D^a \tilde \Gamma_a  )\right]_| \nonumber \\ 
&=& \int d^2 \,  \theta \, \,\overline{s}\,  Tr \left[ c \star (D^a \Gamma_a   ) - \tilde c \star(D^a \tilde \Gamma_a   )  \right]_|.
\end{eqnarray}
In fact in  Landau gauge we can express the sum of the gauge fixing term and the ghost term 
as a combination of  the total BRST and the total anti-BRST variation. Thus in Landau gauge
  sum of the gauge fixing term and the ghost term is given by 
\begin{eqnarray}
\mathcal{L}_{gh} + \mathcal{L}_{gf}  &=& -\frac{1}{2}\int d^2 \,  \theta \, \,s  \overline{s}\,Tr [\Gamma^a \star \Gamma_a - 
\tilde\Gamma^a \star \tilde\Gamma_a]_| \nonumber \\ 
&=& \frac{1}{2}\int d^2 \,  \theta \, \,  \overline{s} s \, Tr [\Gamma^a \star \Gamma_a - \tilde\Gamma^a \star \tilde\Gamma_a]_|.
\end{eqnarray}

\section{ Non-Linear Gauges}
For Yang-Mills theories  in Curci-Ferrari gauge sum of the gauge fixing term and the ghost term can also be expressed 
as a combination of the total BRST and the total anti-BRST variation, 
 for any value of $\alpha$ \cite{n01}. 
In this section we will show that  the sum of the gauge fixing term and the ghost term for this deformed ABJM theory  
in Curci-Ferrari gauge can also be expressed  as a combination of
a total BRST and a total anti-BRST variation, for any value of $\alpha$. 
The BRST transformations  for the deformed ABJM theory  in Curci-Ferrari gauge are given by
\begin{eqnarray}
s \,\Gamma_a = \nabla_a \star c, &&
s \,b  = - [b, c]_\star - [ \overline{c}, [ c,  c]_\star]_\star,\nonumber \\
s \,c  = - [c,  c]_\star,&&
s \,\overline{c}  = b - [\overline{c}, c]_\star, \nonumber \\ 
s \,\tilde \Gamma_a =  \tilde \nabla_a \star \tilde c, &&
s \,\tilde b  = - {[\tilde b, \tilde c]}_\star - [ \tilde{\overline{c}}, [ \tilde c, \tilde c]_\star]_\star \nonumber \\ 
s \,\tilde c = - [\tilde c,  \tilde c]_\star, &&
s \,\tilde{\overline{c}} =\tilde b - [\tilde{\overline{c}}, \tilde c]_\star, \nonumber \\ 
s \, X^{I }  = i(c\star X^{I } -  X^{I }\star \tilde c),  &&
s \, X^{I \dagger } = i(  \tilde c \star  X^{I \dagger }-   X^{I \dagger }\star 
c ),
\nonumber \\
s \, Y^{I } = i(  \tilde{{c}} \star Y^{I }- Y^{I } \star {c}), 
&& s \, Y^{I \dagger } = 
 i({c} \star  Y^{I \dagger } - Y^{I \dagger }\star   \tilde{{c}}).
\end{eqnarray}
The anti-BRST transformation  for this theory in Curci-Ferrari gauge are given by
\begin{eqnarray}
\overline{s}\, \Gamma_a  = \nabla_a \star\overline{c}, &&
\overline{s} \,b  = - [b, \overline{c}]_\star +  [c,[\overline{c},\overline{c}]_\star]_\star,\nonumber \\
\overline{s} \,\overline{c}  = - [\overline{c} , \overline{c}]_\star,&&
\overline{s} \,c  = - b - [\overline{c}, c]_\star, \nonumber \\ 
\overline{s}\, \tilde  \Gamma_a  =\tilde \nabla_a \star \tilde{\overline{c}}, &&
\overline{s} \,\tilde b = - [\tilde b, \tilde{\overline{c}}]_\star +  [\tilde c,[\tilde{\overline{c}},\tilde{\overline{c}}]_\star]_\star,
\nonumber \\ 
\overline{s} \,\tilde{\overline{c}}  = - [\tilde{\overline{c}} , \tilde{\overline{c}}]_\star, &&
\overline{s} \,\tilde c  =- \tilde b - [\tilde{\overline{c}}, \tilde c]_\star, \nonumber \\ 
\overline{s} \, X^{I }  = i(\overline{c}\star X^{I } - X^{I }\star \tilde{\overline{c}}), &&
  \overline{s} \, X^{I \dagger } =
 i(  \tilde{\overline{c}}\star  X^{I \dagger }-  X^{I \dagger }\star \overline{c}),
\nonumber \\
\overline{s} \, Y^{I } = i( \tilde{\overline{c}} \star Y^{I }- Y^{I } \star \overline{c} ), 
&& \overline{s} \, Y^{I \dagger } = 
 i(\overline{c}\star  Y^{I \dagger } - Y^{I \dagger }\star  \tilde{\overline{c}}).
\end{eqnarray}
Both these sets of transformations are also nilpotent. 
\begin{equation}
[s, s]_\star = [\overline{s}, \overline{s}]_\star  = 0. 
\end{equation}
In fact they also satisfy $[s, \overline{s}]_\star = 0$. 
We can now write sum of the gauge fixing term and the ghost term for this deformed ABJM theory
 as a combination of a total BRST and a total  anti-BRST variation, as
\begin{eqnarray}
\mathcal{L}_{gh} + \mathcal{L}_{gf} &=& \frac{1}{2}\int d^2 \,  \theta \, \,s\overline{s} \, Tr  \left[\Gamma^a \star \Gamma_a - \tilde\Gamma^a \star \tilde\Gamma_a 
-  \alpha \overline{c} \star c +  \alpha \tilde{\overline{c}}\star \tilde{c} \right]_| \nonumber \\ 
&=&-\frac{1}{2} \int d^2 \,  \theta \, \,\overline{s} s \, Tr \left[\Gamma^a \star \Gamma_a - \tilde\Gamma^a \star \tilde\Gamma_a
 -  \alpha  \overline{c} \star c_a \right. \nonumber \\
&& \left.\,\,\,\,\,\, \,\,\,\,\,\, \,\,\,\,\,\,\,\,\,\,
 \,\,\,\,\,\, \,\,\,\,\,\, \,\,\,\,\,\,\,\,\,\,
+ \alpha \tilde{\overline{c}}\star \tilde{c} \right]_|.
\end{eqnarray}
In Yang-Mills theory the effect of the addition of a  bare mass to the sum of 
gauge fixing term and the ghost term has been analysed \cite{n001}. 
We  can also generalise  Curci-Ferrari gauge in the deformed ABJM theory  to 
the massive Curci-Ferrari gauge by the addition of a similar bare mass term. 
Thus we can also write the massive Curci-Ferrari type of Lagrangian density for deformed ABJM theory, as 
 \begin{eqnarray}
\mathcal{L}_{gh} + \mathcal{L}_{gf}&=& -\frac{1}{2}\int d^2 \,  \theta \, \,[\overline{s}s+im^2]\, Tr \left[\Gamma^a \star \Gamma_a - \tilde\Gamma^a \star \tilde\Gamma_a 
-  \alpha \overline{c} \star c +  \alpha \tilde{\overline{c}}\star \tilde{c} \right]_| \nonumber \\ 
&=& \frac{1}{2}\int d^2 \,  \theta \, \,[s\overline{s} - im^2]\, Tr \left[\Gamma^a \star \Gamma_a - \tilde\Gamma^a \star \tilde\Gamma_a 
-  \alpha \overline{c} \star c 
\right. \nonumber \\ && \left.\,\,\,\,\,\, \,\,\,\,\,\, \,\,\,\,\,\,\,\,\,\,
 \,\,\,\,\,\, \,\,\,\,\,\, \,\,\,\,\,\,\,\,\,\,
\,\,\,\,\,\, \,\,\,\, \,\,\,\,\,
+ \alpha \tilde{\overline{c}}\star \tilde{c} \right]_|.
\end{eqnarray}
The BRST transformations for the deformed ABJM theory    in this  massive Curci-Ferrari gauge are given by
\begin{eqnarray}
s \,\Gamma_a  = \nabla_a \star c, &&
s \,b  = im^2 c - [b, c]_\star - [ \overline{c}, [ c,  c]_\star]_\star,\nonumber \\
s \,c  = - [c  ,c]_\star, &&
s \,\overline{c}  = b - [\overline{c}, c]_\star, \nonumber \\ 
s \,\tilde \Gamma_a =  \tilde \nabla_a \star \tilde c, &&
s \,\tilde b  =im^2 \tilde{c}- {[\tilde b, \tilde c]}_\star - [ \tilde{\overline{c}}, [ \tilde c, \tilde c]_\star]_\star \nonumber \\ 
s \,\tilde c =- [\tilde c,  \tilde c]_\star, &&
s \,\tilde{\overline{c}} = \tilde b - [\tilde{\overline{c}}, \tilde c]_\star, \nonumber \\ 
s \, X^{I }   =i(c\star X^{I } - X^{I }\star \tilde c),  &&
s \, X^{I \dagger } = i(\tilde c \star  X^{I \dagger }-  X^{I \dagger }
\star  c ),
\nonumber \\
s \, Y^{I } = i( \tilde{{c}}\star Y^{I } - Y^{I }\star {c}) , 
&& s \, Y^{I \dagger } = 
 i({c}\star  Y^{I \dagger } -Y^{I \dagger } \star  \tilde{{c}}). 
\end{eqnarray}
Similarly  the anti-BRST transformation  for the deformed ABJM theory    in this  massive Curci-Ferrari gauge are given by
\begin{eqnarray}
\overline{s}\, \Gamma_a  = \nabla_a \star\overline{c},&&
\overline{s} \,b  = im^2 \overline{c} - [b, \overline{c}]_\star +  [c,[\overline{c},\overline{c}]_\star]_\star,\nonumber \\
\overline{s} \,\overline{c}  = - [\overline{c} , \overline{c}]_\star, &&
\overline{s} \,c = - b - [\overline{c}, c]_\star, \nonumber \\ 
\overline{s}\, \tilde  \Gamma_a  =\tilde \nabla_a \star \tilde{\overline{c}}, &&
\overline{s} \,\tilde b  = im^2 \tilde{\overline{c}} - [\tilde b, \tilde{\overline{c}}]_\star + 
                           [\tilde c,[\tilde{\overline{c}},\tilde{\overline{c}}]_\star]_\star,
                          \nonumber \\ 
\overline{s} \,\tilde{\overline{c}}   = - [\tilde{\overline{c}} , \tilde{\overline{c}}]_\star, &&
\overline{s} \,\tilde c  = - \tilde b - [\tilde{\overline{c}}, \tilde c]_\star, \nonumber \\ 
\overline{s} \, X^{I }  = i(\overline{c}\star X^{I } -  X^{I }\star \tilde{\overline{c}}), &&
  \overline{s} \, X^{I \dagger } =
 i(  \tilde{\overline{c}}\star  X^{I \dagger } - X^{I \dagger } \star \overline{c} ),
\nonumber \\
\overline{s} \, Y^{I } = i(  \tilde{\overline{c}} \star Y^{I }
 -   Y^{I }\star \overline{c})
, 
&& \overline{s} \, Y^{I \dagger } = 
 i(\overline{c}\star Y^{I \dagger } -Y^{I \dagger } \star \tilde{\overline{c}}).
\end{eqnarray}
The addition of bare mass term breaks the nilpotency  of 
the BRST and the anti-BRST transformations. 
The BRST and the anti-BRST transformations now satisfy
\begin{equation}
[s, s]_\star = [\overline{s}, \overline{s}]_\star  \sim 2i m^2.
\end{equation}
However,  in the zero mass limit, 
the nilpotency of the BRST and the anti-BRST transformations is restored. 

\section{Nakanishi-Ojima Algebra}
In Yang-Mills theory it is known that when ever the sum of the  
gauge fixing term and the ghost term can be written as 
a combination of the total BRST and the total anti-BRST variation, 
the total Lagrangian density is invariant under  a set of symmetry 
transformations which obey 
$SL(2, R)$ algebra
\cite{n01}.  Now  for the deformed ABJM theory in 
in the Landau and non-linear gauges, the 
sum of gauge fixing term and ghost term is expressed as a combination of the total BRST and the total anti-BRST variation, 
so we expect the total Lagrangian density for this
 deformed ABJM theory will also be invariant under a set of symmetry 
transformations which obey 
$SL(2, R)$ algebra.  
In fact in these gauges the deformed  ABJM 
theory is also invariant under the following transformations, 
\begin{eqnarray}
 \delta_{1}\, b =  [ c,  c]_\star, 
& \delta_{1}\, \tilde b = 
  [ \tilde c, \tilde c ]_\star, 
& \delta_{1}\, c = 0, 
\nonumber \\
  \delta_{1}\, \tilde c =  0,
  &\delta_{1}\, \overline{c} =  c,
 & \delta_{1}\,  \tilde{\overline{c}} =  \tilde{c}, 
\nonumber \\
 \delta_{1}\, \Gamma_a = 0, 
 & \delta_{1}\,  \tilde{\Gamma}_a =   0,
 & \delta_{1}\, X^{I } = 0, 
\nonumber \\ 
  \delta_{1}\,  X^{I \dagger} =   0,
& \delta_{2}\, b =   [ \overline c,  \overline c]_\star, 
  &\delta_{2}\, \tilde b =   
[ \tilde{\overline{c}}, \tilde{\overline{c}}]_\star, 
\nonumber \\
 \delta_{2}\, c =  \overline{c}, 
&  \delta_{2}\, \tilde c = 
 \tilde{\overline{c}}, 
 & \delta_{2}\, \overline{c} =  0,
\nonumber \\ 
  \delta_{2}\,  \tilde{\overline{c}} =  0,
& \delta_{2}\, X^{I } =  0  
& \delta_{2}\,  X^{I \dagger} =   0.
\end{eqnarray}
In the Landau and Curci-Ferrari gauges these 
 transformations, the BRST transformation and the anti-BRST transformation 
 along with the $FP$-conjugation  form the 
Nakanishi-Ojima $SL(2, R)$ algebra,
\begin{eqnarray}
 [s,s]_{\star} =0, && [\overline{s},\overline{s}]_{\star} =0, \nonumber \\ 
{[s, \overline{s}]}_\star =0, && [\delta_{1}, \delta_{2}]_\star = - 2 \delta_{FP} \nonumber \\ 
{[\delta_{1}, \delta_{FP}]}_\star = -4 \delta_{1}, && [\delta_{2}, \delta_{FP}]_\star = 4 \delta_{2},\nonumber \\ 
{[s, \delta_{FP}]}_\star = - 2s , && [\overline s, \delta_{FP}]_\star = 2\overline s,\nonumber \\ 
{[s, \delta_{1}]}_\star = 0, && [\overline{s}, \delta_{1}]_\star = -2 s,\nonumber \\ 
{[s, \delta_{2}]}_\star = 2 \overline{s}, && [\overline{s}, \delta_{2}]_\star = 0.
\end{eqnarray}
This algebra gets modified due to the presence of the bare mass term in the massive Curci-Ferrari gauge.
This is because the nilpotency of both the BRST and the anti-BRST transformations is broken by the addition 
of a bare mass term. However, even though the nilpotency of the BRST and the anti-BRST transformations broken, the 
$FP$-conjugation is not broken in the massive  Curci-Ferrari gauge. Thus we are  able construct a algebra for the set of 
symmetric transformations  in the massive Curci-Ferrari gauge. 
This  algebra for the set of symmetric transformations  in the 
  massive Curci-Ferrari gauge is given by  
\begin{eqnarray}
 [s,s]_{\star} =-2im^2 \delta_{1}, && [\overline{s},\overline{s}]_{\star} =2im^2 \delta_{2}, \nonumber \\ 
{[s, \overline{s}]}_\star =2im^2 \delta_{FP}, && [\delta_{1}, \delta_{2}]_\star = - 2 \delta_{FP} \nonumber \\ 
{[\delta_{1}, \delta_{FP}]}_\star = -4 \delta_{1}, && [\delta_{2}, \delta_{FP}]_\star = 4 \delta_{2},\nonumber \\ 
{[s, \delta_{FP}]}_\star = - 2s , && [\overline s, \delta_{FP}]_\star = 2\overline s,\nonumber \\ 
{[s, \delta_{1}]}_\star = 0, && [\overline{s}, \delta_{1}]_\star = -2 s,\nonumber \\ 
{[s, \delta_{2}]}_\star = 2 \overline{s}, && [\overline{s}, \delta_{2}]_\star = 0.
\end{eqnarray}
\section{Conserved Charges}

In conventional commutative field theories 
for every symmetry under which the Lagrangian 
density is invariant there is 
a conserved charge obtained from a divergenceless  current
associated with that symmetry of the theory.
   In noncommutative field theories
 even though the variation of the action vanishes for all local
parameters of transformation, 
the divergence of the current need not vanish.
However,  for conventional noncommutative field theories the 
divergence of the current is equal to the Moyal bracket of some
 functions  \cite{msb}. This
Moyal bracket vanishes  for the space-like noncommutativity 
when we integrate on the
continuity equation over all spatial coordinates in 
order to obtain the time variation of the
charge \cite{msb1}. Consequently, the charge associated 
to a symmetry transformation commutes
 with the
Hamiltonian of the theory in this case. 
A similar result will hold for star bracket 
if we are again restricted to spacelike noncommutativity. 
Here again the charge associated 
with a symmetry  transformation will commutes
 with the
Hamiltonian of the theory.
So from now on we shall 
 be  restricted to discussions of
spacelike noncommutativity. So for two local functions
 $X$ and $Z$ associated 
with a symmetry, the divergence of 
the current will be given by
\begin{equation}
 [X, Z]_{\star } = 2 \mathcal{D}^\mu\star J_\mu, 
\end{equation}
where $\mathcal{D}^\mu$ is the ordinary covariant derivative.
As we have restricted
 the discussion to spacelike noncommutativity, we get  
\begin{equation}
 \int d^3y\, \, [X, Z]_{\star } =0.
\end{equation}
The conserved charge is given by 
\begin{equation}
Q = \int d^3 y \, \,   J^0.
\end{equation}

The currents 
associated with the noncommutative  BRST symmetry $J_{(B)}^{\mu}$ and the
noncommutative anti-BRST symmetry $\overline{J}_{(B)}^\mu$  are given by
\begin{eqnarray}
2J_{(B)}^\mu & = &\int d^2 \,  \theta \, \, Tr 
\left[ \frac{ \partial L_{eff}  }{\partial \mathcal{D}_{\mu} \Gamma_{b} } 
\star  s\, 
\Gamma_{b} +
 \frac{ \partial L_{eff}  }{\partial  \mathcal{D}_{\mu} c } \star  s\, c
   +
\frac{ \partial L_{eff}  }{\partial  \mathcal{D}_{\mu} \overline{c} }
 \star  s\, \overline{c} \right. \nonumber \\&&
 \,\,\,\,\,\, \,\,\,\,\,\,\, \,\,\,\,\,\, \,\,\,\,\,\,\,\,\,\, +
\frac{ \partial L_{eff}  }{\partial \mathcal{D}_{\mu} b } \star   s\, b
 +
\frac{ \partial L_{eff}  }{\partial \mathcal{D}_{\mu} \tilde \Gamma_{a} } \star  s\, 
\tilde \Gamma_{a} +
 \frac{ \partial L_{eff}  }{\partial  \mathcal{D}_{\mu} \tilde c }
 \star  s\, \tilde 
c
 \nonumber \\&& 
 \,\,\,\,\,\, \,\,\,\,\,\,\, \,\,\,\,\,\, \,\,\,\,\,\,\,\,\,\, +
\frac{ \partial L_{eff}  }{\partial  \mathcal{D}_{\mu} \tilde{\overline{c}} }
 \star  s\, \tilde{\overline{c}} + 
\frac{ \partial L_{eff}  }{\partial \mathcal{D}_{\mu} \tilde b } \star  
 s\,\tilde
  b  + \frac{ \partial L_{eff}  }{\partial \mathcal{D}_{\mu}  X^I } \star  
 s\,
  X^I 
 \nonumber \\&&
 \,\,\,\,\,\, \,\,\,\,\,\,\, \,\,\,\,\,\, \,\,\,\,\,\,\,\,\,\, +
\frac{ \partial L_{eff}  }{\partial  \mathcal{D}_{\mu} X^{I \dagger } }
 \star  s\, X^{I \dagger } + 
\frac{ \partial L_{eff}  }{\partial \mathcal{D}_{\mu} Y^I } \star  
 s\,Y^I
 \nonumber \\&&\left. 
 \,\,\,\,\,\, \,\,\,\,\,\,\, \,\,\,\,\,\, \,\,\,\,\,\,\,\,\,\,
 + \frac{ \partial L_{eff}  }{\partial \mathcal{D}_{\mu}  Y^{I \dagger } } \star  
 s\,
  Y^{I \dagger } \right]_|,
\nonumber \\
 2\overline{J}_{(B)}^\mu  & = &\int d^2 \,  \theta \, \, Tr 
\left[ \frac{ \partial L_{eff}  }{\partial \mathcal{D}_{\mu} \Gamma_{b} }
 \star \overline{s}\, 
\Gamma_{b} +
 \frac{ \partial L_{eff}  }{\partial  \mathcal{D}_{\mu} c } \star  \overline{s}\, c
  +
\frac{ \partial L_{eff}  }{\partial  \mathcal{D}_{\mu} \overline{c} }
 \star  \overline{s}\, \overline{c} \right. \nonumber \\&& 
 \,\,\,\,\,\, \,\,\,\,\,\,\, \,\,\,\,\,\, \,\,\,\,\,\,\,\,\,\,+ 
\frac{ \partial L_{eff}  }{\partial \mathcal{D}_{\mu} b } \star 
  \overline{s}\, b
 +
\frac{ \partial L_{eff}  }{\partial \mathcal{D}_{\mu} \tilde \Gamma_{a} } 
\star  \overline{s}\, 
\tilde \Gamma_{a}  +
 \frac{ \partial L_{eff}  }{\partial  \mathcal{D}_{\mu} \tilde c } \star  \overline{s}
\, \tilde 
c 
 \nonumber \\&&
 \,\,\,\,\,\, \,\,\,\,\,\,\, \,\,\,\,\,\, \,\,\,\,\,\,\,\,\,\, +
\frac{ \partial L_{eff}  }{\partial  \mathcal{D}_{\mu} \tilde{\overline{c}} }
 \star  \overline{s}\, \tilde{\overline{c}} + 
\frac{ \partial L_{eff}  }{\partial \mathcal{D}_{\mu} \tilde b } \star  
 \overline{s} \,\tilde 
  b
+ \frac{ \partial L_{eff}  }{\partial \mathcal{D}_{\mu}  X^I } \star  
  \overline{s}\,
  X^I 
 \nonumber \\&&
 \,\,\,\,\,\, \,\,\,\,\,\,\, \,\,\,\,\,\, \,\,\,\,\,\,\,\,\,\, +
\frac{ \partial L_{eff}  }{\partial  \mathcal{D}_{\mu} X^{I \dagger } }
 \star   \overline{s}\, X^{I \dagger } + 
\frac{ \partial L_{eff}  }{\partial \mathcal{D}_{\mu} Y^I } \star  
  \overline{s}\,Y^I 
 \nonumber \\&&\left. 
 \,\,\,\,\,\, \,\,\,\,\,\,\, \,\,\,\,\,\, \,\,\,\,\,\,\,\,\,\,
+ \frac{ \partial L_{eff}  }{\partial \mathcal{D}_{\mu}  Y^{I \dagger } } \star  
  \overline{s}\,
 Y^{I \dagger } 
\right]_|,
\end{eqnarray}
where
\begin{equation}
 \int d^2 \,  \theta \, \, [L_{eff}]_| = \mathcal{L}_c + \mathcal{L}_{gh}
 + \mathcal{L}_{gf}.
\end{equation}
Hence, the BRST charge  $Q_B$ and the anti-BRST charge $\overline{Q}_B$ 
associated with the currents $J_{(B)}^\mu$ and 
$\overline{J}_{(B)}^\mu$ are conserved, 
\begin{eqnarray}
 Q_B &=& \int d^3 y  \, \,  J_{(B)}^0 , \nonumber \\ 
 \overline{Q}_B &=& \int d^3 y  \, \,  \overline{J}_{(B)}^0.
\end{eqnarray}
The  BRST charge  $Q_B$ and the anti-BRST charge $\overline{Q}_B$
 are both nilpotent for all gauges except the massive Curci-Ferrari gauge, 
\begin{equation}
 Q_B^2 =  \overline{Q}_B^2 =0. 
\end{equation}
However, for massive Curci-Ferrari gauge these charges are not nilpotent 
\begin{eqnarray}
  Q_B^2 &\neq& 0, \nonumber \\
\overline{Q}_B^2 &\neq& 0. 
\end{eqnarray}
The nilpotency of $Q_B$ and $\overline{Q}_B$ is very 
important to isolate the physical Hilbert space 
and prove the unitarity of the $S$-matrix.
 This is what will be done in the next section. 

\section{Physical Sub-Space} 
The total Lagrangian  which is formed by the sum of
 the original Lagrangian, the gauge fixing 
term and the ghost term is invariant under 
the noncommutative BRST and the noncommutative anti-BRST transformations.  
As the charges $Q_B$ and $\overline{Q}_B$  are nilpotent for all gauges except the  massive Curci-Ferrari gauge,
 so  their action on any field 
 twice will vanish for all gauges except the  massive Curci-Ferrari gauge. 
So for any state $|\phi\rangle$ in a gauge other than the  massive Curci-Ferrari gauge, we have 
\begin{eqnarray}
 Q_B^2 |\phi\rangle &=& 0, \nonumber \\ 
 \overline{Q}_B^2 |\phi\rangle &=& 0.
\end{eqnarray}
We shall now restrict out discussion to gauges other than the  massive Curci-Ferrari gauge. 
The  physical states  $ |\phi_p \rangle $ can now be defined as  states that are annihilated by $Q_B$ 
\begin{equation}
 Q_B |\phi_p \rangle =0. 
\end{equation}
We can also  define the physical states as states that are annihilated by $\overline{Q}_B$
 \begin{equation}
 \overline{Q}_B |\phi_p \rangle =0. 
\end{equation}
We will obtain the  same result by using any of these as the definition for the physical sates. 
Now as we 
get the same physical result by using either the noncommutative  BRST or the noncommutative  anti-BRST charge,
 we will denote them both by $Q$, so
$Q$ represents both   $Q_B$ and $\overline{Q}_B$.
Thus the physical states $Q |\phi_p \rangle $  are annihilated by $Q$,
\begin{equation}
 Q |\phi_p \rangle =0. 
\end{equation}

This criterion divides the Fock space into three parts, $\mathcal{H}_0, \mathcal{H}_1$ and $\mathcal{H}_2$.  
The space $\mathcal{H}_1$, comprises of those  states that are not annihilated by $Q$. 
So if $|\phi_1\rangle$ is any state in $\mathcal{H}_1$, then we have
\begin{equation}
 Q | \phi_1\rangle \neq 0.
\end{equation}
The space $\mathcal{H}_2$ comprises of those states that are obtained by the action of $Q$ on states
 belonging to $\mathcal{H}_1$. So if  $|\phi_2\rangle$ is any state in $\mathcal{H}_2$, then we have
\begin{equation}
 | \phi_2\rangle = Q | \phi_1\rangle.
\end{equation}
Thus we have 
\begin{equation}
 Q| \phi_2\rangle = Q^2 | \phi_1\rangle =0. 
\end{equation}
So all the states in $\mathcal{H}_2$ are annihilated by $Q$. 
The space $\mathcal{H}_0$ comprises of those states that are annihilated by $Q$ and are 
not obtained by the action of $Q$ on any state belonging to $\mathcal{H}_1$. So if  $|\phi_0\rangle$ is any state in $\mathcal{H}_0$,
 then we have
\begin{eqnarray}
 Q | \phi_0\rangle &=&0, \\ 
 | \phi_0\rangle &\neq& Q | \phi_1\rangle.
\end{eqnarray}
Clearly the physical states $|\phi_p \rangle $ can only belong to $\mathcal{H}_0$ or $\mathcal{H}_2$.
This is because any  state in   $\mathcal{H}_0$ or $\mathcal{H}_2$ is  annihilated by $Q$. 
However, any state in $\mathcal{H}_2$ will be orthogonal to all  physical states including itself. 
\begin{eqnarray}
  \langle \phi_p |\phi_2\rangle& =&  \langle \phi_p|( Q |\phi_1\rangle)\nonumber \\ &=& ( \langle \phi_p| Q) |\phi_1\rangle =0.
\end{eqnarray}
Thus two physical states  that differ from each other by a state in $\mathcal{H}_2$  will be indistinguishable,
\begin{equation}
 |\phi_p\rangle = |\phi_p \rangle +  | \phi_2 \rangle.
\end{equation}
So all the relevant physical states actually lie in $\mathcal{H}_0$.

Now if the asymptotic physical states are given by 
\begin{eqnarray}
 |\phi_{pa,out}\rangle &=& |\phi_{pa}, t \to \infty\rangle, \nonumber \\
 |\phi_{pb,in}\rangle &=& |\phi_{pb}, t \to- \infty\rangle,
\end{eqnarray}
 then a typical $\mathcal{S}$-matrix element can be written as
\begin{equation}
\langle\phi_{pa,out}|\phi_{pb,in}\rangle = \langle\phi_{pa}|\mathcal{S}^{\dagger}\mathcal{S}|\phi_{pb}\rangle.
\end{equation}
Now as the noncommutative BRST and the noncommutative anti-BRST charges  
are conserved charges, so they commute with the Hamiltonian and thus the time evolution of any physical state will 
also be annihilated by  $Q$, 
\begin{equation}
 Q \mathcal{S} |\phi_{pb}\rangle =0.
\end{equation}  
This implies that the states $\mathcal{S}|\phi_{pb}\rangle$ must be a linear combination of states in $\mathcal{H}_0$ and $\mathcal{H}_2$. 
However, as the states in  $\mathcal{H}_2$ have zero inner product with one another and also with states in $\mathcal{H}_0$, so the 
only contributions come from states in $\mathcal{H}_0$. So we can write 
\begin{equation}
\langle\phi_{pa}|\mathcal{S}^{\dagger}\mathcal{S}|\phi_{pb}\rangle = \sum_{i}\langle\phi_{pa}|\mathcal{S}^{\dagger}|\phi_{0,i}\rangle
\langle\phi_{0,i}| \mathcal{S}|\phi_{pb}\rangle.
\end{equation}
Since the full $\mathcal{S}$-matrix is unitary this relation implies that the  $S$-matrix restricted to physical sub-space is also unitarity. 
It may be noted that the nilpotency of the noncommutative BRST and the noncommutative anti-BRST charges was essential for 
proving the unitarity of the resultant theory. Now as the noncommutative BRST and the noncommutative anti-BRST 
charges are not nilpotent  in the massive Curci-Ferrari  gauge,
\begin{eqnarray}
 Q_B^2 |\phi\rangle &\neq& 0, \nonumber \\ 
 \overline{Q}_B^2 |\phi\rangle &\neq& 0,
\end{eqnarray}
so the $\mathcal{S}$ does not factorise in the massive Curci-Ferrari  gauge
\begin{equation}
 \langle\phi_{pa}|\mathcal{S}^{\dagger}\mathcal{S}|\phi_{pb}\rangle \neq \sum_{i}\langle\phi_{pa}|\mathcal{S}^{\dagger}|\phi_{0,i}\rangle
\langle\phi_{0,i}| \mathcal{S}|\phi_{pb}\rangle,
\end{equation}
and thus  the resultant theory is not unitarity.
However, even though  this noncommutative deformation
 is not unitary in the massive Curci-Ferrari  gauge, 
 the nilpotency is restored 
in the zero mass limit. Thus the unitarity is 
also restored in the zero mass limit. 
\section{Conclusion}
In this paper we studied a noncommutative deformation of the ABJM theory 
in $N=1$ superspace formalism. 
In performing our analyses the noncommutative fields were related to 
ordinary ones and the product of these noncommutative fields was related 
to a star product of ordinary fields. The main focus of the paper was 
to generalize  some results that are known  for Yang-Mills theories 
to this deformed ABJM theory. 
So we 
 analysed the behaviour of the BRST 
and the anti-BRST symmetries for  this deformed  ABJM  theory,
 is linear as well as non-linear gauges.
  We  have expressed 
the sum of gauge fixing term and the ghost term  for this deformed
  ABJM  theory 
as a combination of the total BRST and the total anti-BRST variation, 
in the  Landau gauge.
Furthermore, this was achieved  for an arbitrary value of $\alpha$ 
by the making the BRST and the anti-BRST transformations  non-linear.
The addition of a bare mass term  
violated the nilpotency of the BRST and the
 anti-BRST transformations and 
this in turn breaks the unitarity of  the theory. 
We have also  shown that in 
Landau and Curci-Ferrari gauges the  
deformed ABJM  theory is invariant under 
 Nakanishi-Ojima $SL(2, R)$ algebra. 
We also have also analysed the effect  
that the addition of  a bare mass term has on this algebra.

In Yang-Mills theories the presence of non-linear terms gives rise  to an effective
potential whose vacuum configuration favours the formation of off-diagonal ghost condensates \cite{gc1}.
The ghost condensation in Yang-Mills theories also occurs in the Landau gauge \cite{gc4}.
 The ghost condensation in Yang-Mills theories breaks the $SL(2,R)$ symmetry which exists in these gauges. 
It will be interesting to investigate if the ghost
 condensation  in this deformed
 ABJM theory also leads to a dynamic breaking 
of $SL(2,R)$ symmetry. 

The infinite temporal derivatives  occur 
in the product of fields for this noncommutative ABJM theory  due to 
$B^{0\mu}$ and $A^{0 a }$. This will give rise to non-local behaviour
  in the deformed ABJM theory. 
This in general will lead to a violation of the unitarity 
of the deformed ABJM theory. 
However, if we restrict the deformation
 of the ABJM theory to spacelike noncommutativity i.e., we set 
$B^{0\mu} = A^{0a} =0$, then this problems will not occur.
 It will then be possible to construct the Norther's charger's 
corresponding to the BRST and the anti-BRST symmetries and 
use them to project out the physical states. 
 It will be interesting to construct these BRST and  anti-BRST charger's for this theory 
and use them to find the physical states in this theory.  As the nilpotency of the BRST and 
the anti-BRST transformations  is violated in the massive Curci-Ferrari gauge, 
so we expect that unitarity will also be violated in that gauge, even after restricting to spacelike noncommutativity

 ABJM theory has    been used to study various examples of $AdS_4/CFT_3$ correspondence \cite{6}-\cite{10}. 
In fact $AdS_4/CFT_3$  has also been used to analyse   fractional quantum Hall effect \cite{e}.  
Fractional quantum Hall effect in ABJM theory has also been analysed \cite{e1}. 
In ABJM theory   $D6$-branes wrapped over $AdS_4 \times S^3/Z_2$ 
in type $IIA$ super-string theory on $AdS_4 \times  CP^3$ give its dual description with $N=3$ supersymmetry. 
 In the presence of fractional branes, the ABJM theory can model the fractional 
quantum Hall effect, with $RR$-fields regarded as the external 
electric-magnetic field. In this model 
  addition of the flavour $D6$-brane describes a class of fractional quantum Hall plateau transition. 
It will be interesting to analyse 
fractional quantum Hall effect, with $RR$-fields regarded as the external 
electric-magnetic field in the deformed superspace. We can expect that 
addition of the flavour $D6$-brane might describes
 a class of fractional quantum Hall plateau transition
in the deformed superspace ABJM theory also.


\begin{thebibliography}{99}
\bibitem{1}A. Gustavsson, JHEP. 0804, 083 (2008)
 \bibitem{2}J. Bagger and N. Lambert, JHEP. 0802, 105 (2008)
\bibitem{3}J. Bagger and N. Lambert, Phys. Rev. D77, 065008 (2008)
 \bibitem{4}M. A. Bandres, A. E. Lipstein and J. H. Schwarz, JHEP. 0809, 027 (2008)
\bibitem{5}E. Antonyan, A. A. Tseytlin, Phys. Rev. D79, 046002 (2009)
\bibitem{su}
O. Aharony, O. Bergman, D. L. Jafferis and J. Maldacena, JHEP. 0810,  091 (2008)
\bibitem{apjm}
O-Kab Kwon, P. Oh and  J. Sohn, JHEP.  0908, 093 (2009)
\bibitem{14}A. Mauri and  A. C. Petkou, Phys. Lett. B666,  527 (2008)
\bibitem{ab1}
S. V. Ketov, S. Kobayashi, Phys. Rev. D83, 045003 (2011) 
 \bibitem{sw} N. Seiberg and E. Witten, 
 JHEP. 9909, 032 (1999)
\bibitem{dn}
M. R. Douglas and N. A. Nekrasov,  Rev.  Mod.  Phys. 73, 977 (2001)
\bibitem{dfr} S. Doplicher, K. Fredenhagen and  J. E. Roberts,
    Commun. Math. Phys. 172, 187 (1995)
\bibitem{co} A. Connes,  Non commutative geometry. Academic
Press, Inc. London (1990)
\bibitem{la} G. Landi,  An introduction to noncommutative
spaces and their geometries. Springer-Verlag (1997).
\bibitem{mssw}
J. Madore, S. Schraml, P. Schupp and J. Wess,  Eur.  Phys. J.  C16, 161 (2000)
\bibitem{bcz}
D. Brace, B. L. Cerchiai and B. Zumino,  Int. J. Mod. Phys. A
1751, 205 (2002)
\bibitem{av}
L. Alvarez-Gaume and M. A. Vazquez-Mozo, Nucl. Phys. B668, 293 (2003)
\bibitem{ov} H. Ooguri and C. Vafa, 	Adv. Theor. Math. Phys. 7, 53 (2003)
\bibitem{se} N. Seiberg, JHEP.  0306, 010 (2003)
\bibitem{bgn} J. de Boer, P. A. Grassi and P. van Nieuwenhuizen,
 Phys. Lett. B574,  98 (2003)
\bibitem{beta1}
E. Chang-Young, H.Kim and H. Nakajima, JHEP. 0804, 004 (2008) 
\bibitem{beta2}
K. Araki, T.  Inami, H. Nakajima and Y. Saito, JHEP. 0601, 109 (2006) 
\bibitem{beta3}
J. S. Cook, J. Math. Phys. 47, 012304  (2006)
\bibitem{beta4}
Y. Kobayashi and S. Sasaki, Phys.Rev. D72, 065015  (2005) 
\bibitem{bs} N. Berkovits and N. Seiberg, JHEP. 0307, 010 (2003)
\bibitem{14om}
C. S. Chu and E. Sezgin,  JHEP. 9712, 001
(1997) 
\bibitem{41s}
P. S. Howe and E. Sezgin,  Phys. Lett. B 394, 62 (1997) 
\bibitem{41ss}
M. Aganagic, J. Park, C. Popescu and J. H. Schwarz,  Nucl. Phys. B 496, 191 (1997)
\bibitem{41ssa}
P. Pasti, D. P. Sorokin and M. Tonin,  Phys. Lett. B 398, 41 (1997) 
\bibitem{41ssb}
I. A. Bandos, K. Lechner, A. Nurmagambetov, P. Pasti, D. P. Sorokin and M. Tonin,
 Phys. Rev. Lett. 78, 4332 (1997)
\bibitem{42s}
P. S. Howe, E. Sezgin and P. C. West, Phys. Lett. B 399, 49 (1997)
\bibitem{cssw}
N. Grandi and G.A. Silva, Phys. Lett. B 507,  345 (2001)
\bibitem{cssw1}
 M. Picariello, A. Quadri and S. P. Sorella, 
JHEP. 0201, 045 (2002) 
\bibitem{abcd}
N. Nakanishi and I. Ojima, Covariant operator formalism of gauge theories and quantum gravity - World Sci. Lect. Notes. Phys. (1990)
 \bibitem{16}J. Fjelstad and S. Hwang, Phys. Lett. B466, 227 (1999)
\bibitem{n01}
H. Nicolai, Phys. Lett.  B89, 341 (1980)
\bibitem{n06}
D. Dudal, V. E. R. Lemes, M. S. Sarandy, S. P. Sorella, and M. Picariello.
 JHEP. 12, 008  (2002)
\bibitem{n09}
R. Marnelius and U. Quaade, J. Math. Phys. 36, 3289 (1995)
\bibitem{n04}
N. Nakanishi and I. Ojima, Zeit. Phys. C6, 155  (1980)
\bibitem{n001}
G. Curci and R. Ferrari, B 63, 91  (1976)
\bibitem{17}M. Chaichian and W. F. Chen, Z.Y. Zhu, Phys. Lett. B387, 785 (1996)
\bibitem{18}L.P. Colatto, M.A. De Andrade, O.M. Del Cima, D.H.T. Franco, J.A.
Helayel-Neto and O. Piguet, J. Phys. G24, 1301 (1998)
\bibitem{19}C. P. Constantinidis, O. Piguet and W. Spalenza, Eur. Phys. J. C33, 443
(2004)
\bibitem{20g}
A. Das and M. M. S. Jabbari, JHEP. 0106, 028 (2001) 
\bibitem{21g}
A. A. Bichl, J. M. Grimstrup, V. Putz and  M. Schweda, JHEP.
 0007, 046  (2000) 
\bibitem{k}
E. Witten, Commun. Math. Phys. 121, 351 (1989)
\bibitem{gc1}
Kei-Ichi Kondo and Toru Shinohara,  Phys. Lett. B 491, 263 (2000)
\bibitem{gc4}
R. Delbourgo and P. D. Jarvis, 
J. Phys. A15, 611  (1982)
\bibitem{6}I. R. Klebanov and A. M. Polyakov, Phys. Lett. B550, 213 (2002)
\bibitem{7}J. H. Schwarz, JHEP. 0411, 078 (2004)
 \bibitem{8}C. Ahn, H. Kim, B. H. Lee and H. S. Yang, Phys. Rev. D61, 066002
(2000)
 \bibitem{9}B. Chen and J. B. Wu, JHEP. 096, 0809 ( 2008)
 \bibitem{10}M. Benna, I. Klebanov, T. Klose and M. Smedback, JHEP. 0809, 072
(2008)
\bibitem{e}
M. Fujita, W. Li, S. Ryu and T. Takayanagi,  JHEP. 0906, 066 (2009)
\bibitem{e1}
Y. Hikida, Wei Li and T. Takayanagi, JHEP. 0907, 065 (2009) 
 \bibitem{msb}
A. Micu and M. M. S. Jabbari,  JHEP. $\bf{0101}$, 025 (2001)
\bibitem{msb1}
M. Soroush, Phys. Rev. $\bf{D 67}$, 105005  (2003) 
\end{thebibliography}
\end{document}